# Coordination of resources at the edge of the electricity grid: systematic review and taxonomy

F. Charbonnier[a,*], T. Morstyn[b] and M. McCulloch[a]

[a]Department of Engineering Science, University of Oxford, UK
[b]School of Engineering, University of Edinburgh, UK

ABSTRACT

This paper proposes a novel taxonomy of coordination strategies for distributed energy resources at the edge of the electricity grid, based on a systematic analysis of key literature trends. The coordination of distributed energy resources such as decentralised generation and flexibility sources is critical for decarbonising electricity and achieving climate goals. The literature on the topic is growing exponentially; however, there is ambiguity in the terminology used to date. We seek to resolve this lack of clarity by synthesising the categories of coordination strategies in a novel exhaustive, mutually exclusive taxonomy based on agency, information and game type. The relevance of these concepts in the literature is illustrated through a systematic literature review of 84,741 publications using a structured topic search query. Then 93 selected coordination strategies are analysed in more detail and mapped onto this framework. Clarity on structural assumptions is key for selecting appropriate coordination strategies for differing contexts within energy systems. We argue that a plurality of complementary strategies is needed to coordinate energy systems' different components and achieve deep decarbonisation.

## List of Abbreviations

| | | | |
|---|---|---|---|
| ADMM | Alternating direction of multipliers | ICT | Information and control technology |
| DLT | Distributed ledge technology | L1 | First layer of the proposed taxonomy |
| DER | Distributed energy resource | L2 | Second layer of the proposed taxonomy |
| DLMP | Distribution locational marginal pricing | L3 | Third layer of the proposed taxonomy |
| DR | Demand response | ML | Machine learning |
| DSO | Distribution system operator | PV | photovoltaic |
| EMS | Energy management system | P2P | peer-to-peer |
| EV | Electric vehicle | RL | Reinforcement learning |
| GDPR | General data protection regulation | TOU | Time of use |
| HVAC | Heating, ventilation, and air conditioning | UK | United Kingdom |

Table 1: Nomenclature

## 1. Introduction

This paper investigates how strategies for the coordination of grid-edge energy resources – distributed energy resources (DERs) connected at the distribution network level [1] – can be synthesised in a hierarchical classification according to their structural similarities and differences, referred to as a taxonomy [2].

Coordination of grid-edge resources could make a major contribution to power system decarbonisation, which is critical for keeping anthropogenic warming below 1.5°C above pre-industrial levels. Exceeding this limit would severely increase the risks of extreme climate events and associated impacts on health, livelihood, security and economic growth. Widespread electrification of primary energy provision and decarbonisation of the power sector are therefore necessary, and renewable power supplies could be required to supply 70% to 85% of electricity globally by 2050[1] [3]. While numerous energy generation and enabling storage and flexibility technologies exist and are emerging, a key challenge is their integration at an unprecedented scale. The coordination of energy flexibility in all sectors is needed for the integration of high levels of intermittent renewable energy, from the transmission to the distribution network [3].

*Corresponding author
flora.charbonnier@eng.ox.ac.uk (F. Charbonnier)
ORCID(s): 0000-0003-3174-0362 (F. Charbonnier); 0000-0003-2781-9588 (T. Morstyn); 0000-0001-5378-1128 (M. McCulloch)

To achieve these coordination objectives, a robust decarbonised power system will rely on two major structural features: decentralisation and demand response (DR) [4]. As the flexibility, control, and data ownership are increasingly decentralised, DERs can provide such decentralised DR, as well as operational services such as frequency regulation, spinning reserves provision, voltage management, system balancing and network congestion management [5, 6]. Moreover, the coordination of flexible DERs can yield numerous local and global co-benefits: reduction of environmental concerns, reduced energy transport and storage costs, improved grid stability, alignment of peak demand with decarbonised energy provision in time and space, reduced costs of peaking plants and capacity reserves, deferral of transmission and distribution grid upgrades, energy independence and security, incentives for the contribution of numerous actors in investment, reduced bills for consumers, and enhanced social cohesion [7–10]. Extensive research is therefore conducted to extend the realm of coordination to small, grid-edge resources in the distribution grid, and the scholarship on the coordination of resources at the edge of the electricity grid has been growing exponentially since 1995 (Figure 1). Particularly, novel research is seeking to tackle the challenges of computation and control at the scale of millions of units [11], privacy concerns and acceptability issues [7, 8, 12], as well as increased uncertainty at the local level [13].

Although numerous reviews focus on specific areas within the field of grid-edge energy resources coordination, they concentrate on limited aspects rather than systematically reviewing the landscape of coordination strategies, and use conflicting terminology. Some reviews focus on one type of grid-edge resource only, for example on residential thermal energy storage [14] or deferrable loads [15]. Others focus on specific segments of the electricity grid, such as on the residential context [16, 17], or on microgrids [18–21]. Previous reviews have analysed particular coordination methods such as market-based coordination [5, 6, 9, 22–26], optimisation [27], particle swarm optimisation and genetic algorithms [28] and reinforcement learning (RL) [7]. Others have investigated specific technological tools enabling coordination, such as distributed ledger technologies (DLT) [29] and digital tools such as modelling, simulation and hierarchical control [30]. Tohidi et al. review grid-edge resources coordination indirectly, by investigating the possible interactions between local and central electricity markets [31]. Guerrero et al. investigate the technical issues associated with the implementation of behind-the-meter DERs coordination strategies in a low-voltage network [32]. Despite all these individual thematic reviews, there is no systematic review and taxonomy of the field of coordination of resources at the edge of the electricity grid with applicability across energy technologies. The terminology used to categorise DER coordination strategies within the literature have overlaps and inconsistencies, with terms such as "peer-to-peer", "multi-agent" or "transactive energy" which may refer to coordination frameworks with fundamentally different structural features. The resulting unproductive linguistic ambiguity impedes effective communication and understanding in the field, which can hinder both academic progress and collaboration with industry.

This paper seeks to bridge this gap to bring greater clarity to the classification and terminology in the field of DER coordination. The principal contributions of this paper are:

- The development of a novel comprehensive taxonomy for distributed resources coordination strategies, which aims at clarifying the structural features of coordination strategies, as the terminology currently used to describe them is often ambiguous.
- The identification of key research themes corresponding to the coordination categories through a systematic review of the literature relevant to the coordination of distributed resources at the edge of the electricity grid.
- The analysis of 93 coordination strategies mapped to the taxonomy through a detailed literature review to illustrate the wealth of coordination strategies corresponding to each category of the taxonomy.

The rest of this paper is organised as follows. In Section 2, a novel synthesised taxonomy of coordination strategies is developed based on the types of agency, information flow structure and game type. We analyse the relevance of this categorisation in key associated research trends identified by a systematic literature review. In Section 3, we demonstrate the ambiguity of some of the current terminology used in the field, which is not consistently aligned with

[1] interquartile range in 1.5°C pathways with no or limited overshoot (high confidence)

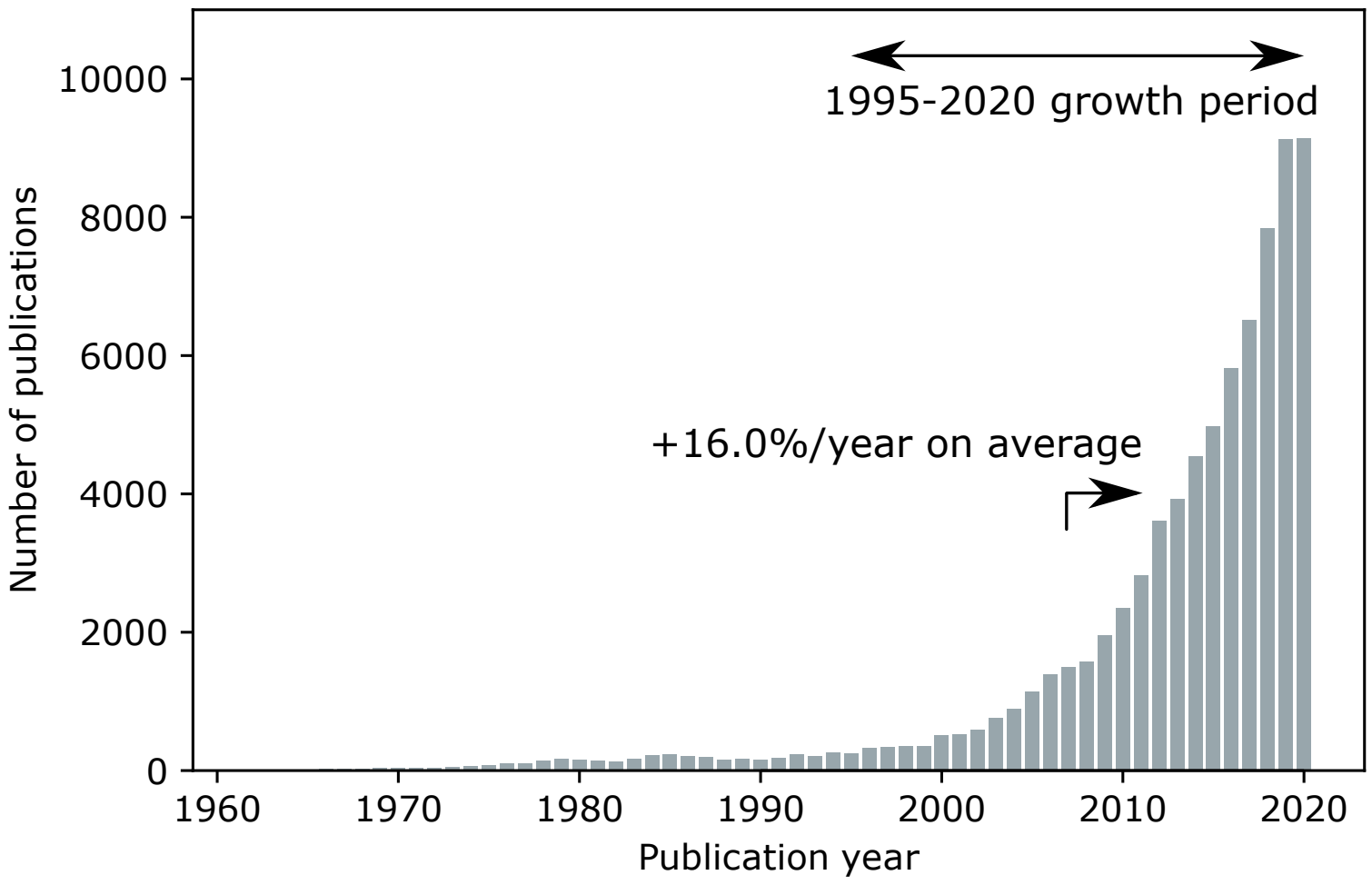

**Figure 1:** Number of publications in a selected body of literature on the coordination of electricity grid-edge resources between 1962 and 2020. See Appendix A.1 for the details of the systematic literature search. The number of publications grew exponentially by 16.0% per year on average between 1995 and 2020.

the structural features of coordination categories, and highlight how the taxonomy helps resolve this ambiguity. We then provide a more detailed review to clarify the boundaries of the coordination categories by mapping 93 selected recent coordination strategies onto the taxonomy. In Section 4, we look at potential applications of the taxonomy for coordination strategy selection. We argue based on the evidence provided that context-specific heterogeneous complementary strategies are needed to coordinate different flexibility sources in energy systems and suggest specific paths forward for research such as residential energy coordination. Finally, we conclude in Section 5.

## 2. Taxonomy of distributed energy resources coordination strategies

This section develops a novel taxonomy for the categories of DER coordination based on objective structural features of the individual strategies. The taxonomy clarifies the interplays between agency decentralisation, information flows, and game motivations. We illustrate the relevance of these categories by analysing major research themes in a systematic literature review of the scholarship on grid-edge energy resources coordination.

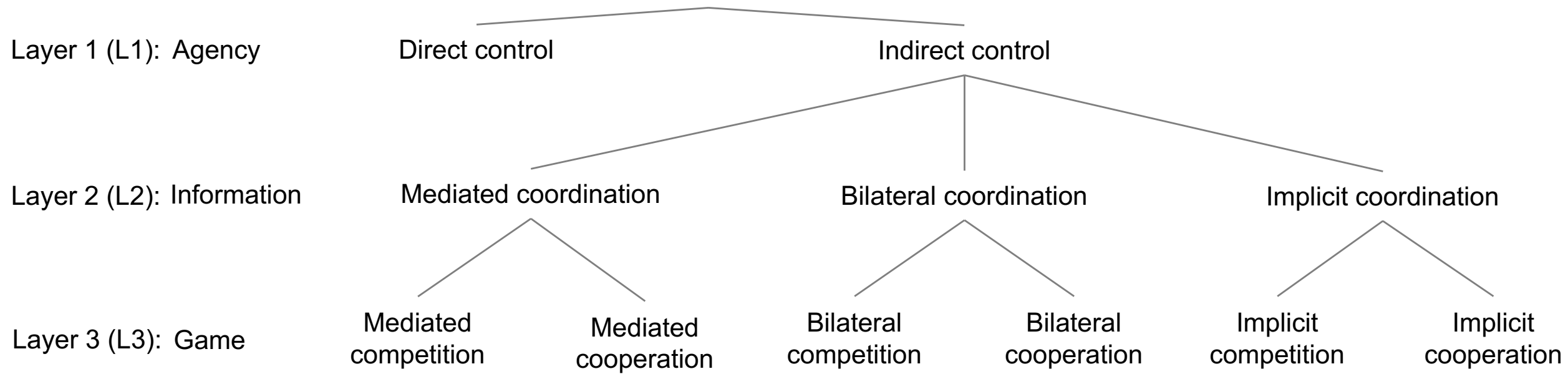

**Figure 2:** Systematic taxonomy of categories of coordination, based on the answer to three questions corresponding to the three layers of the classification: (L1) Agency: Are coordinated units operated independently? (L2) Information: How is individual information shared? (L3) Game type: Do units compete or cooperate?

As shown in Figure 2, three questions corresponding to the three layers of the taxonomy L1, L2 and L3 are used to systematically classify any coordination strategy into exhaustive and mutually exclusive structural categories: (L1)

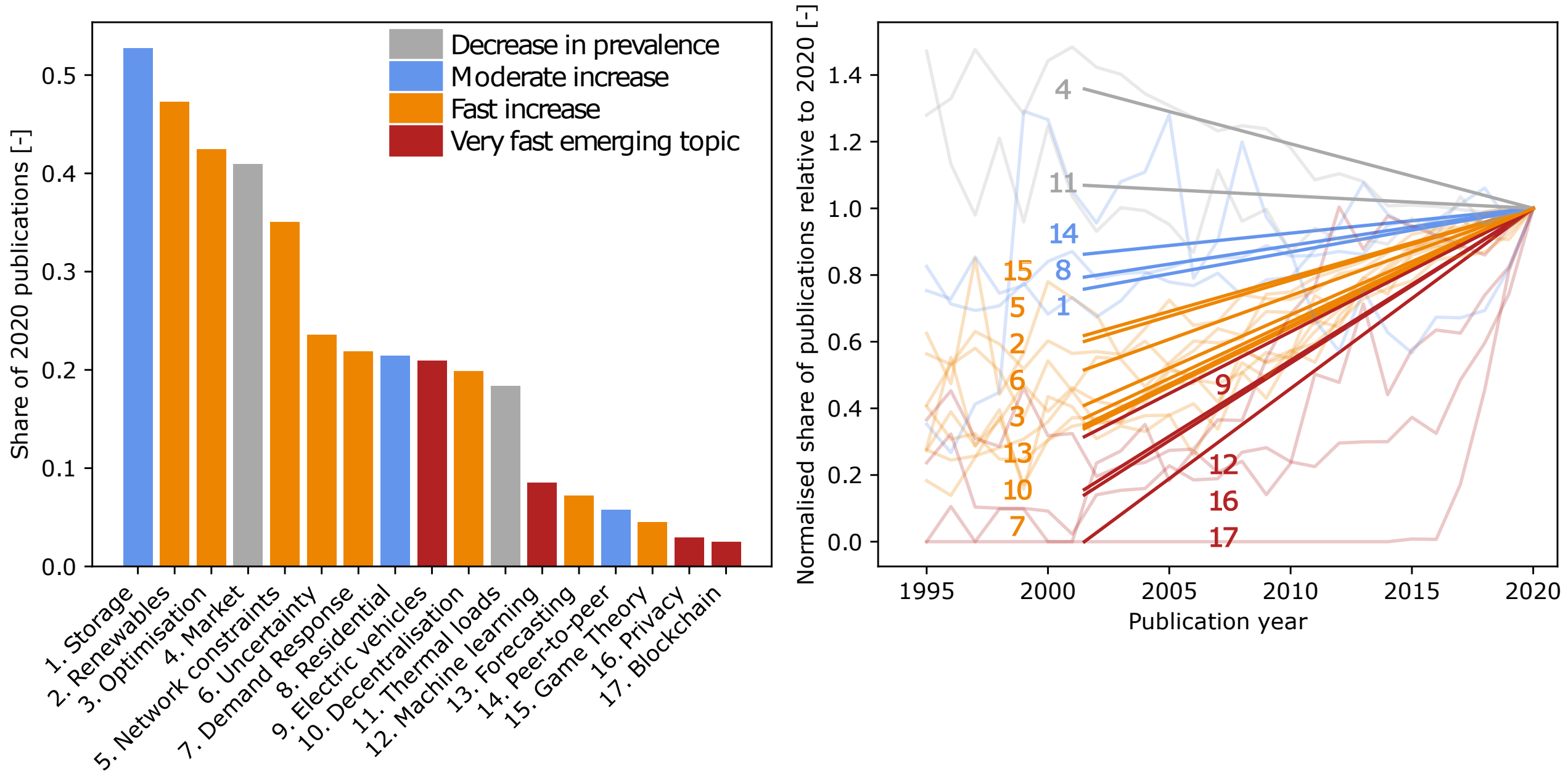

**Figure 3:** The left-hand side plot shows the prevalence of research themes in 2020, i.e. the share of all publications on the coordination of grid-edge electricity resources that include theme-related keywords in their titles and abstracts. Faint lines on the right-hand side show the prevalence of themes over time normalised by that in 2020. Opaque lines show the change between the average normalised prevalences of themes in the first half of the period (1995-2007) and that in 2020. Moderate, fast, and very fast increases correspond to average prevalences in the first half of less than 100%, two thirds and one third of that in 2020 respectively. See Appendix A.2 for the details of the theme identification. Note that all themes have increased in absolute numbers over the time period as the total scholarship under study has increased; this figure analyses the relative importance of individual themes within the scholarship.

Agency: Are coordinated units operated independently? (L2) Information: How is individual information shared? (L3) Game type: Do units compete or cooperate? In the rest of this section, we present these three elucidating questions in more detail and illustrate their relevance in key associated research themes identified in a systematic literature review of grid-edge energy resource coordination. A structured topic search query in the Scopus search engine [33] was conducted, selecting literature that lies at the intersection of the concepts of coordination, grid-edge participation and electric resources (see Appendix A). Key research themes were identified within this body of literature using keywords from titles and abstracts. The absolute and relative prevalence of these major themes are displayed on Figure 3, so that trends corresponding to the main layers of the taxonomy may be identified.

## 2.1. Agency: Are coordinated units operated independently?

We call *direct control* the case where a central entity has full access to the information from all units and can decide on their control actions. A system objective is pursued regardless of whether this is beneficial for single units. In *indirect control*, prosumers make decisions autonomously at the local level.

In this paper, we define a unit as one or multiple energy resources providing flexibility in operation in the same location, whose assets and information are operated by the same entity. The proposed taxonomy is agnostic to the type of coordinated unit, and each structural category is applicable across a wide range of applications. Given the increasing urgency of climate change and consequently the need for electricity decarbonisation, there has been a rapid increase in interest in the coordination of both renewable energy assets (47.2% prevalence in the literature on grid-edge energy resource coordination in 2020 in Figure 3) and of key associated enabling flexibility resources. The integration of large shares of intermittent renewable electricity generation will require flexibility to align the power generated with consumption and to provide operational services [34]. Words relating to storage thus occurred in just over half (52.7%) of the relevant literature. While storage traditionally takes the form of standalone stationary batteries, a very fast emerging number of publications are investigating the coordination of electric vehicle (EV) batteries. Moreover, other elements of flexibility may also be conceptualised as virtual storage, such as that provided by demand response [35, 36], which is also a fast emerging research theme. Coordinating thermal loads has for example long been studied in this body of literature though there is now a slight relative decline in interest. On the other hand, residential flexibility resources receive growing attention due to the outstanding DR potential provided by increasing electrification of household loads and ownership of smart resources. However, there are significant challenges to their coordination due to the large number of small units and the required interaction with user needs, which prevent direct monitoring and controllability of distributed units.

Given these trends, the question of agency is critical as direct controllability of resources from different owners with different objectives and resources is challenging [37]. As a result, interest in research themes that place independent preferences and decisions at the heart of coordination strategies such as peer-to-peer trading, game theory and blockchain structures has been increasing, though they are so far only mentioned in 5.7%, 4.5% and 2.5% of the relevant literature respectively (see Figure 3). The choice of *direct* or *indirect control* depends on the level of intelligence and flexibility of individual units, the alignment of individual interests, as well as other contextual legal and physical constraints [18].

### 2.2. Information: How is individual information shared?

*Indirect control* strategies can in turn be subdivided between *mediated* coordination where a central entity collects information[2] about prosumers, *bilateral* coordination where prosumers only communicate information bilaterally with one another, and *implicit coordination* where personal information is not shared, with at most one-way communication of market information to prosumers without feedback of personal information.

The information structure of coordination strategies is increasingly relevant as interest in the decentralisation of energy resources is rapidly increasing in the literature (see Figure 3), transforming the way data is owned and communicated. Recently developed strategies have sought to complement the existing centralised control by fully utilising not only the generation and flexibility offered locally but also the distributed data ownership, computation power and communication capabilities. While exchanging information provides value, it also raises numerous ethical, trust and very fast increasing privacy concerns (see Figure 3). Data gathered may be improperly used, both information directly collected from prosumers and other sensitive information about users' habits and lifestyles inferred from their behaviour and interactions [38]. As such, a trade-off exists between the value of sharing information and both the degree of privacy [39] and the costs of communication and control infrastructure.

With decentralisation and limited local data availability comes increasing uncertainty, which is a fast increasing research theme (23.5% prevalence in 2020), especially as the share of intermittent and unpredictable electricity-generating technologies increases. When both physical resources and ownership of data are distributed, complete and instantaneous information flows for determining adequate controls cannot be obtained. An optimal control signal based on inaccurate information yields suboptimal outcomes, with potential negative impacts on physical systems such as the electricity grid. While some techniques such as robust and least-regret optimisation account for this uncertainty, the fields of forecasting (7.2% prevalence in this corpus in 2020 in Figure 3) and machine learning (ML) (8.5%) are of fast and very fast increasing interest respectively. Depending on the information structure, these can help bridge a lack of data at the local level. Forecasting aims at reducing uncertainty in predictions, although renewable resources availability and behaviour-influenced demand are inherently unpredictable, especially at the local scale. In ML, agents learn incrementally from collected experience and are able to take statistically optimal actions with incomplete information within an uncertain, stochastic environment. The use of ML for DER coordination is facilitated by the increasing availability of data measurement in the electricity grid and by reduced computational requirements compared to physical-based models [30].

Data availability and communication is therefore key in determining which type of coordination strategy to use.

### 2.3. Game type: Do units compete or cooperate?

Finally, we classify the coordination strategies based on the type of game prosumers are playing.

Prosumers may willingly *cooperate* towards the maximisation of common global objectives, such that some additional social value may be obtained relative to the sum of individual utility if decisions were made in isolation [40]. In certain situations, agents could perform actions with huge benefits for society as a whole, but would derive insufficient personal benefits from bringing these positive externalities in a competitive framework [36]. Regulatory intervention and cooperation between various actors are therefore needed to unlock these benefits. A common objective may simply be the equitable achievement of all private self-interested objectives, or may be wider in scope. A common objective may for example be managing network constraints. Although the challenge of maintaining grid operation within acceptable physical limits has always been prevalent in distributed electricity resources coordination, the share of publications dealing with power distribution networks management has been steadily increasing due to the aforementioned decentralisation and uncertainty of resources (Figure 3) [9]. Concerns include voltage fluctuations and imbalance, current harmonics, network congestion and stability issues [30], especially in the distribution grid where independent assets are not readily controlled, and at the interface between transmission and distribution grid. Failing to account for such network constraints in coordination strategies may lead to negative impacts during operation and to increased investment costs [32].

[2] Information refers to load and generation curve predictions, bids or constraints, for example.

Alternatively, prosumers may *compete* with one another, seeking to maximise their own local utility[3] only, whether purely based on profits or taking into account personal preferences. This market-based control aligns loosely with definitions of "transactive energy" with "users [...] considered as self-interested with heterogenous preferences" [32] where "price signal plays a key role, since it is a universal language for all type of devices and systems for making a decision and performing the optimal usage of the resources" [9]. Note that although individual prosumers aim to maximise their own utility only, the overall market mechanism may be designed to direct those self-interested decisions towards additional objectives such as network management or aggregator profits to obtain collective outcomes as close as possible to an optimisation result [4]. Electricity markets particularly differ from markets in other sectors in that they need a designer and a system operator, who may solve an optimisation problem for optimal dispatch. Therefore, the result of trading can be modelled as an optimisation problem, although no optimisation actually occurs. Ideal market signals are analogous to dual variables in optimisation problems for marginal pricing frameworks [43]. As an example, long-term climate change mitigation may be incentivised using marginal climate costs pricing signals to align personal interests to global ones [44]. However, in the absence of the inclusion of externalities in prices (greenhouse gas emissions, network constraints management, grid losses, network utilisation), markets may not maximise the welfare of the entire system. Optimal real-time operation of energy resources is therefore an interdisciplinary matter, where both optimisation and market-based approaches can co-exist and complement each other – they were each mentioned in just above 40% of the selected body of literature in 2020 (Figure 3). Market-based control of electricity assets has been extensively researched concurrently to the wave of liberalisation of the system in the 1990s, while the research focus is now increasingly on optimisation strategies as the reach of information and communication technologies (ICT) is extending to the edge of the grid (Figure 3).

## 3. Detailed literature review

In this section, we conduct a detailed review of the literature on the coordination of grid-edge electricity resources through the lens of the taxonomy developed in Section 2. Firstly, we demonstrate the ambiguity of some of the current terminology used in the literature. Then, we seek to resolve this ambiguity by demonstrating how all strategies can be mapped onto the taxonomy categories, using a sample of ninety-three coordination strategy definitions from the literature.

### 3.1. The ambiguous terminology of control architectures

Using the taxonomy developed in Section 2, we find that the terminology used in the literature does not point to clear structural features of coordination strategies. Linguistic ambiguity in the field hinders the clarity of structural assumptions about control paradigms, with terms being used to denote inconsistent meanings.

We take as examples the terms "multi-agent", "peer-to-peer" and "transactive energy". As shown in Table 2 and in the bullet points below, each of these labels has been given to such varied fundamental approaches to coordination that they have lost their specificity.

- Strategies have been labelled as multi-agent systems in contexts often departing from the established definition proposed by Wooldridge (2002) [45]: "An agent is a computer system that is capable of independent action on behalf of its user or owner. In other words, an agent can figure out for itself what it needs to do in order to satisfy its design objectives, rather than having to be told explicitly what to do at any given moment. A multi-agent system consists of a number of agents, which interact with one another, typically by exchanging messages through some computer network infrastructure". For example, a framework was described as a "multi-agent system" although distributed units had no agency, and residential agents were directly controlled centrally. [46] While there is no right or wrong definition, the various meanings tied to the term has prevented this labelling from conveying clear specific meaning on its own.

[3] Utility was defined by Jeremy Bentham as "that property in any object, whereby it tends to produce benefit, advantage, pleasure, good, or happiness (all this in the present case comes to the same thing) or (what comes again to the same thing) to prevent the happening of mischief, pain, evil, or unhappiness to the party whose interest is considered" [41]. A utility function in turn is an economist's convenient representation of an individual's preferences that permits mathematical analysis [42].

| Coordination category | "multi-agent" labelling | "peer-to-peer" labelling | "transactive energy " labelling |
|---|---|---|---|
| Direct control | 46 | | 32, 50 |
| Mediated competition | 51–55 | 56–58 | 26, 32, 50, 58–61 |
| Mediated cooperation | 62–64 | 47, 65 | 11 |
| Bilateral competition | | 26, 56, 66–70 | 32, 61, 68 |
| Bilateral cooperation | 19, 71 | 47, 48 | 32 |
| Implicit competition | | | 32, 50, 61 |
| Implicit cooperation | 72 | | |

**Table 2**
Example uses of the labels "multi-agent", "peer-to-peer" and "transactive energy" in different structural coordination categories across the proposed taxonomy. The terminology therefore loses specificity.

- Similarly, Table 3 illustrates the widely varying levels of specificity with which "peer-to-peer" systems are defined. Moreover, publications in turn describe P2P mechanisms as inherently cooperative [40, 47] and competitive [48].
- While most frameworks labelled as "transactive energy" refer to competitive frameworks, this is inconsistently used. Thus, in one source, transactive energy systems were defined as using "automated device bidding [...] through market processes" [49], while in another distributed optimisation techniques have been framed at transactive energy, with all agents cooperating to reach a single objective by decomposing the problem at the user level [32].

The wealth of concepts and definitions appearing in the literature testifies to the rapidly increasing interest the field is attracting. However, the use of overlapping terminology may be counter-productive as we lose the ability to label individual coordination structures clearly and precisely. The taxonomy developed in Section 2 may help resolve this ambiguity and improve communication by providing objective structural criteria for classification. As the body of literature concerned with the coordination of resources at the edge of the electricity grid is growing exponentially and extends to every layer of our electricity systems, precise terminology is needed for relevant problems to begin to be understood and assessed.

We now demonstrate how the taxonomy can be used to synthesise the terminology used in the literature by mapping 93 strategies and elements of terminology onto the coordination categories. We illustrate the wealth of strategies corresponding to each of these categories and clarify their structural differences and similarities.

## 3.2. Direct Control

*Direct control* is identified in the top layer (L1) of the taxonomy in Figure 2. Units forfeit their data and control to a central entity. Different terms in the literature align with this category of control architecture, such as "centralised control" [19], "direct load control" [13], and "centralised dispatch" [70].

Direct control may be technically accomplished in different ways. Receding horizon global optimisation frameworks is commonly proposed [78–81], where the uncertainty of optimisation inputs can be accounted for in with stochastic optimisation [79] and by subjecting the scheduler to the worst-case photovoltaic (PV) predictions [81] among others. The amount of data shared by units to allow the centralised entity to make control decisions may vary from "top-down switching" where the entity uses statistics to directly turn on and off loads without consumer information given, to "centralized optimization" where the central entity truly can optimise the search space thanks to full access to information [50]. Design and operation of the system may be considered together under a total system optimisation [82]. Schedulers using RL are proposed that directly controls residential appliances [83]. Direct control may also be rule-based, for example to synchronise fridges' thermal storages [84] or to design some direct-load control architecture for an aggregator based on priority stacks [85] or other heuristics [46]. In event-based direct control, customers receive incentive payments for allowing the utility a degree of control over certain equipment; the utility can reduce the loads in response to a variety of trigger conditions such as grid conditions or system temperature [10].

| Reference | Peer-to-peer definition |
|---|---|
| 69, 73 | The trading between suppliers and consumers |
| 32 | Users trading energy among themselves with a limited or no intervention of a third party |
| 9, 58, 74 | The sharing of resources at the edge of the network, at the distribution level |
| 56, 75 | Selling and buying of surplus electricity for small-scale residential and commercial producers and consumers |
| 68 | The coordination of large number of small-scale producing and consuming units |
| 26, 66 | The ability of individual prosumers to make independent decisions, being able to chose when to trade, how much, and at what price |
| 60, 76 | Markets allowing prosumers to engage in bilateral trades |
| 70, 77 | Mechanisms allowing prosumers to negotiate with one another directly |
| 48 | Mechanisms allowing prosumers to negotiate with one another directly and without any direct influence of a central controller |
| 67 | A network in which members share resources and information to attain energy-related objectives, without any intervention from a third party controller, and where any peer can be added or removed without altering the operational structure of the system |

**Table 3**
In increasing order of specificity, examples of definitions for "peer-to-peer" trading found in the literature. The label therefore no longer points to a specific structural coordination category.

Direct control methods are particularly suited to small-scale microgrids [78], or if the aggregator owns resources directly. However, individual domestic households may not be inclined to give up control of appliances in their own homes, expressing privacy and security concerns. Direct control may moreover pose a significant computation burden or even be computationally intractable at large scale [57]. It could necessitate extensive ICT infrastructure and be highly vulnerable to the single-point-of-failure of the central controller or communication links [9, 18, 30]. Leveraging full information availability and control, typical objectives for direct control strategies are global operation costs minimisations, energy arbitrage, peak shaving, load shifting and the provision of ancillary services (see Table B.1).

## 3.3. Indirect control

In *indirect control* (Figure 2: L1), coordinated units have agency and are operated independently at the local level. This can be either a manual or automated response. Indirect control may be broken down into *mediated*, *bilateral* and *implicit coordination* (Figure 2: L2) as presented below.

### 3.3.1. Mediated coordination

In *mediated coordination*, a central entity collects information about prosumers (Figure 2: L2) and redistributes information back to them, such as matchings between peers, price signals or partial results from a central optimisation for use in local computation. Two-way communication allows the utility to monitor real-time flexibility availability and leads reliably to intended outcomes as information is available to steer individual actions in the desired direction. This category aligns with the concept of "coordinated approach" [32]. Different well established mathematical theories can be used to design mediated coordination, such as auction theory and optimisation theory.

This approach is most suited to situations with infrastructure and acceptance for reliably and safely sharing individual information to a central entity. Prosumers may however have security and privacy concerns over sharing their information centrally. Biased information, due to the inaccuracy of forecasts or to strategic behaviour [77], may moreover lead to inefficient or even infeasible decisions. It is not guaranteed that prosumers will have the will, interest or ability to receive, interpret, and respond according to the centrally computed signals [12]. Indeed, both local ICT infrastructure and prosumer engagement are necessary to utilise flexibility.

In the third layer of the taxonomy (Figure 2: L3), mediated coordination can be broken down into *competitive* and *cooperative* categories as defined below.

- *Mediated competition* involves competitive individual units which maximise their own objective function only. The mediator collects information from units and sends back signals that they believe will incentivise globally optimal action, such as price signals or prosumer matchings. Although participating units are selfish, further objectives may be served depending on the design of the strategy, such as the aggregated provision of ancillary services, the minimisation of global operational costs and the flattening of load curves (see Table B.1).

  For example, the central entity may send unidirectional price signals to customers based on information such as prosumers' costs, constraints and day-ahead forecasts, corresponding to a proposed definition of a "community market" [26]. In a "price-responsive mode", prosumers make decisions in response to market price signals by the operator [61]. Pricing signals corresponding to Distribution Locational Marginal Pricing (DLMP) with import-export price spreads can incentivise adequate prosumer decisions [70]. Pricing signals that are adaptive to centrally monitored consumption levels internalise the externality effect of customers, where an increase in customer demand would create a negative externality effect for other customers' price rate [16]. Given knowledge of previously negotiated prosumer trades, the central distributor can compute and allocate losses using its global knowledge of the distribution network and transactions [59]. RL can inform both the dynamic price signal [55, 86], and the prosumer response to price signals [55, 87]. The exchange of information between prosumers and the mediator may also be iterative – such as in a non-cooperative Stackelberg games where both utilities and consumers try to maximise their utility by iteratively updating their prices and demand respectively [88], and with prosumers learning to compute a Nash equilibrium through iterations of exchange of information between prosumers and operator, with electricity prices dependent upon the aggregate demands at each time step [89].

  Alternatively, another approach to mediated competition uses "organized markets" [10] analogous to wholesale markets, where the central entity collects bids and set trades and match prosumers centrally. The labels "system-centric matching" [56], "coordinated market" [26], "transactive control" [50] and "explicit demand response" [28] fit in this category. Instead of sending control signals directly, bidding approaches promote price discovery, seeking to reveal the marginal value of consumption given fragmented unknown private technical characteristics, personal preferences and opportunity costs of participants. This bidding reveals the information required to obtain control signals without directly using private information such as supply and demand functions used in formal optimisation [49]. Examples of this include a unilateral auction mechanisms [58], continuous double auction mechanisms [32, 49, 57], demand reduction bids where customers send bids consisting of their available demand reduction capacity and their requested price [10], an auction mechanism where the difference between accepted offers and accepted bids are allocated for congestion costs [90], and nested negotiations between DSO and aggregators and between aggregators and prosumers [60]. RL algorithms may be used to facilitate bidding either to refine individual bidding strategies [52, 54, 75, 91, 92] or to dictate the double auction market clearing [51, 93].

- In *mediated cooperation*, the mediator collects information from units (Figure 2: L2) and sends back signals to the prosumers, expecting them to cooperate (L3). The aim is to maximise system-wide welfare by minimising total costs, sometimes against immediate individual preferences or interests. Customers need technical and financial support to install the necessary infrastructure to react accordingly to signals from the utility [10]. The mediator may take on a "social planner" role, a benevolent coordinator who chooses an economic policy either to maximise a social welfare function or to attain a Pareto efficient allocation [94]. Typical applications of mediated cooperation seek to address both prosumer utility maximisation and cooperative objectives such as peak shaving, ancillary services provision or minimisation or operation costs (see Table B.1). Though distributed units work together to achieve common goals in both cases, mediated cooperation differs to direct control in that the mediator does not have full access to personal information for central direct control of units. Instead, only partial personal information is shared so that central signals may guide cooperation without full knowledge of individual utility functions or device states, while final computations and decision-making are on the individual level.

  The mediator's role is to collect and redistribute information from and to prosumers such that they can take actions compatible with global objectives. In a semiautonomous mode of operation [95], instructions on how to respond to locally available grid signals are updated centrally and regularly based on power systems characteristics and sent to local appliances. In an iterative method, the power system operator shares network information to prosumers, who can then privately arrange trades taking into account their marginal losses and constraints, and communicate updated trades to the operator [96]. Based on the new set of trades, the operator then broadcasts updated network information. In other examples of iterations between central and distributed computations, prosumers run local optimisations with augmented terms to align local prosumer coordination with global constraints and goals, and send back results to adapt iteratively centrally computed signals informing, in turn, the local optimisations [73, 97]. In an other strategy, an aggregator collects information on the flexibility of building agents and the distribution network [64]. Each building energy management system (EMS) then aims to meet flexibility requests by the aggregator with minimum discomfort, using a multi-agent Q-learning method that includes an explicit model of the other agents.

In other strategies, prosumers cooperate with the help of a mediator by forming coalitions to achieve their goals. In a "community-based market", members agree on common goals and trade together as a coherent group in markets, the trades being handled by a community manager [47]. This is very similar to the "energy collective" where a supervisory node facilitates the interface to different markets [11]. Prosumer coalitions can be computed centrally using game theory based on their information and the optimal operation is solved centrally for the coalition [40]. In another example, prosumers participate in a coalition formation game with an auctioneer acting as an intermediate, i.e. a mediated competition process precedes the cooperation [65].

RL algorithms have been proposed where agents interacting with a central entity learn to cooperate to pursue common goals. Agents can use RL with individual agent fitness functions that include the objective functions of other agents while satisfying local operation constraints, and learnings are shared within cooperative swarms to rapidly update their knowledge [53]. In other frameworks, the central entity sends each prosumer information about 8 neighbours [62, 63]. Each prosumer learns to pursue common goals for the nine prosumers, such as aligning loads to renewable energy provision and avoiding high total load.

### 3.3.2. Bilateral coordination

In *bilateral coordination*, prosumers only communicate information bilaterally with one another with no central authority or organised pool (Figure 2: L2). This category is most suited to contexts where communication infrastructure is available and where privacy requirements are compatible with the bilateral communication of individual information. Robustness to communication failures increases relative to mediated coordination as there is no longer a centralisation of data with a single point of failure. As the system size increases, the number of communication iterations until algorithm convergence increases, requiring adequate computational resources with limited communication network latency for feasibility [32].

The safe way of implementing distributed transactions to ensure data protection is an ongoing subject of research. Many trial projects use DLTs such as the blockchain, allowing for a decentralised marketplace without an intermediary. An incorruptible secure digital ledger records financial transactions permanently in a decentralised fashion [98], in multiple points without a single point of failure [21]. They however present legal, environmental and implementation risks. Legal risks of non-compliance to regulations such as to the European Union General Data Protection Regulation (GDPR) arise as personal data cannot be erased and stakeholders cannot control their own data, with no clear accountable entity [99]. Furthermore, negative externalities associated with blockchains have been well documented, due to proof-of-work processes requiring ever greater computational power and energy consumption for mining, causing environmental and health impacts contradicting the very aim of decarbonisation [100]. Finally, implementation risks such as scalability concerns, facilitated money laundering, fraud, and tax evasion should be carefully weighed before creating a dependency at a large scale of our electricity systems on blockchain technologies [101].

We now further differentiate the *bilateral competition* and *cooperation* (Figure 2: L3) categories.

- In a *bilateral competition* framework, peers directly negotiate energy transactions with one another (Figure 2: L2) and chose trades to maximise their utility (see Table B.1). This is analogous to "peer-centric" mechanisms [56],"decentralised markets" [26], and "bilateral transactive bids" [61]. The offer of autonomy, expression of individual preferences and market transparency may appeal to prosumers. However, agents may exhibit bounded rationality, unable to know the competitors' production decision and profit functions [102, 103] and to process all the necessary information to perform a cognitively burdensome exhaustive optimisation of rational utility in real decision-making situations. They may often rely on simple rules of thumb or decision heuristics, which may widen the optimality gap [104] and dampen the reaction of agents to market or policy signals [105]. To avoid potentially serious consequences for the physical grid system, markets should therefore be designed to elucidate consumer preferences and appraise their impact on both market outcomes and the physical power networks [47, 103].

Some bilateral competition strategies have been proposed with blockchain-managed bidding mechanisms [66–68]. Others rely on bilateral contract networks, multi-sided matching markets where units form bilateral downstream and upstream contracts to sell outputs and buy inputs [32]. The potential for stability of a bilateral contract network was extensively analysed by Fleiner et al. [106] and further explored by Morstyn et al., who established a fully-distributed, iterative P2P negotiation mechanism [69, 70] .

- In *bilateral cooperation*, prosumers also use bilateral negotiation, but to cooperatively implement theoretically-proven market solutions for the whole system (Figure 2: L3). Prosumers compute local responses taking into account not only their personal preferences and profits but also system objectives and constraints. A range of cooperative objectives may be pursued (see Table B.1). This category closely aligns with the definition of "cooperative control of multi-agent systems" [107], "multi-agent control strategy" [19], and "distributed control" [30].

  Bilateral cooperation strategies can be implemented using iterative negotiation to reach shared objectives [32, 70]. In other proposals, common objective functions are maximised using variations of distributed optimisation such as dual decomposition and the alternating direction method of multipliers (ADMM) [47]. For example, distributed optimisation can be used [48], where global and local constraints are decoupled by applying Lagrangian multipliers. Scalability is however a concern for these methods. In ADMM, although there are established proofs of convergence, the convergence time may be sensitive to problem-specific numerical properties and may be operationally impractical [11]. In other proposed strategies, agents use transfer learning with distributed W-learning to achieve local and system objectives [71].

  Bilateral cooperation may be vulnerable to the risk of strategic behaviour or gaming by market participants, as the convergence to optimal cooperative outcomes depends on information supplied by individual units and on their cooperating rather than trying to maximise their individual utilities only [77]. Computational issues may arise as the complexity increases with the number of DERs at scale [70]. Moreover, cooperative game theory-based profit-sharing allocations that incentivise cooperation in a way that is robust to strategic behaviour is exponentially complex, limiting scalability [40]. Safeguarding measures include ring-fencing the distribution networks, with a clear definition and allocation of distribution costs in incentive regulation [108]. This approach is suitable for systems with larger number of nodes and high complexity, and offers cost reductions for systems that frequently need to be expanded relative to centrally controlled systems with more expensive communication and control infrastructure [30].

#### *3.3.3. Implicit coordination*

In *implicit coordination*, prosumers do not share their personal information with a central entity nor with their peers, but may monitor their current wider information environment (e.g. wholesale prices) or utilise information about past system characteristics to inform their independent decision-making (Figure 2). While bilateral trades are effective coordination tools, the physical reality of the electricity network only recognises injection and extraction points, with electrons flowing independently of the bilateral financial transactions the market participants agree upon. The problem of coordination thus boils down to the sum of individual decisions to import and export electricity. This case corresponds to the definition of "decentralised control" [30] where "the control decisions are made individually at each DER by its local controller using the local information".

Advantages of implicit coordination include reduced complexity and costs of the ICT infrastructure, enhanced privacy, self-control and acceptability for users, robustness against failures, and reliability [30, 32]. The simplicity of implicit coordination reduces implementation failure risks due to the lack of interoperability between different smart grid elements, as connectivity between individual hardware and software components was reported as the most common obstacle reported in real-world programmes [10, 37]. One-way communicating devices are highly cost-effective as they incur lower upfront costs than two-way communication, though they do not allow monitoring and verification of the DR impact with accurate precision [10]. A suboptimality gap will exist as the strategies cannot be informed by real-time personal data from all units, as well as unexpected problems in the case of uncooperative operation [30]. Therefore, these strategies are most suited where the cost of ICT infrastructure for each unit outweighs the potential benefits of communication, particularly for small units where there may be privacy concerns.

We now map coordination strategies onto the *implicit competition* and *implicit cooperation* coordination categories.

- In essence, *Implicit competition* is the status quo scenario. Most consumers today do not share their data and behave competitively (Figure 2: L3), i.e. solely maximise their own utility to serve local goals such as individual costs and comfort (see Table B.1). This paradigm is also called "price-reactive system" [50], "smart pricing" [10], "autonomous mode" [61] and "uncoordinated approach" [32]. Critical-peak pricing, time-of-use pricing (TOU) and real-time pricing are common ways of implementing implicit coordination [10].

  As self-interested units are only concerned with the individual scale, many strategies leave the realm of distributed units coordination to focus on single unit local energy management, for example with load scheduling algorithms under TOU pricing [109]. RL models can help optimise personal utility by exploiting opportunities for energy arbitrage [110, 111], learning optimal appliance scheduling decisions by interacting with user feedback [112], and conserving energy while ensuring user comfort [113]. Rule-based and RL-based control of heating, ventilation, and air conditioning (HVAC) resources have been proposed [114]. The uses of particle swarm optimisation and genetic algorithm have also been investigated to control thermal energy storage [28].

  Self-interested responses to price signals, without factoring in the impact on the whole system of the sum of their individual actions, result in suboptimal system outcomes, especially if deployed at a large scale. For example, a concern is that all loads receive the same incentive, the natural diversity on which the grid relies may be diminished [115], and the peak potentially merely displaced, with overloads on upstream transformers. A case study thus results in capacity issues for high DER penetration if global network constraints and objectives are not taken into account in decision-making [32]. Moreover, uncoordinated reactions to price signals may be difficult to predict without knowledge of devices' states and end users' preferences [50]. Contrary to mediated competition where the coordination signals are dynamically updated based on real-time communication of individual information, in implicit competition the prices are sent unidirectionally, and thus need to be carefully selected ahead of operation. The market mechanism should be designed such that self-interested schedules add up to limit suboptimality.

- In the *implicit cooperation* category, units do not share information (Figure 2: L2) but make individual decisions cooperatively to optimise global objectives (L3).

  Operating without either direct, centralised control or bilateral sharing of information has been proposed, where prosumers cooperate to reach system-wide benefits statistically. Within this category are found the "decentralised control strategies" [19] and "autonomous control" [30], which focus on voltage and droop control [20], a line frequency control method based on local information only. Fully autonomous modes of control have been defined to offer a hard-wired primary response to grid frequency deviations [95].

  This paper makes the argument that the space of possible control and coordination strategies and methods that would fall under this categorisation is under-researched, with very few strategies exploring implicit coordination beyond frequency control (see Table B.1). Two such examples were identified [72, 116], where agents use RL algorithms without communicating personal data and seek to statistically assess the impact of their actions on achieving common objectives.

The mapping of coordination strategies onto the proposed taxonomy presented throughout this section is synthesised in Table 5 and extended in Table B.1. These tables illustrate both the wealth of possible strategies within each category and the lack of specificity of the terminology which is used across different structural categories. In Table B.1, the type of objectives pursued is classified to aid in the selection of adequate strategies given contextual structural constraints and objectives. As presented in this section, multiple direct control strategies provide load shifting, peak shaving and ancillary services, while competitive strategies tend to focus on individual prosumer utility. Typical cooperative strategies seek to minimise operation costs while considering prosumer utility. Multiple different types of controlled DER units are considered in each category (generation, storage, thermal loads, other flexible loads). Strategies that consider network losses and constraints are also identified, as network management was previously identified as a key challenge for the coordination of DERs.

| Coordination category | Example key words for strategy description | Refs. |
|---|---|---|
| Direct control | "centralized", "centralised dispatch", "direct load control", "event-based DR", "model predictive control", "Operator instructions", "optimal power flow", "optimization", "top-down switching" | 10, 19, 30, 32, 46, 50, 61, 70, 78–85 |
| Mediated competition | "adaptive consumption-level pricing scheme", "bidding strategy", "coordinated market", "demand response", "double auction", "dynamic pricing algorithm", "incentive-based demand response", "indirect customer-to-customer energy trading", "market-based control", "non-cooperative", "organized markets", "price-responsive", "P2P market","Stackelberg game", "transactive control", "transactive energy", "unidirectional pricing" | 10, 16, 26, 28, 32, 49–52, 54–61, 67, 70, 75, 86–93 |
| Mediated cooperation | "community-based", "coordinated multilateral trades", "distributed demand response", "distributed optimization", "distribution locational marginal costs and hierarchical decomposition", "energy collectives", "joint action learning", "prosumer coalitions", "P2P trading", "transfer learning" | 11, 40, 47, 53, 62, 63, 65, 73, 95–97 |
| Bilateral competition | "auction", "bilateral contracts", "bilateral transactive bids", "blockchain", "decentralized market", "distributed", "local energy market", "matching theory", "P2P trading" | 26, 32, 65, 66, 68–70, 106 |
| Bilateral cooperation | "bilateral energy trading", "decentralized", "distributed","distributed dispatch", "parallel", "P2P" | 19, 30, 32, 47, 48, 70, 71 |
| Implicit competition | "autonomous", "energy arbitrage", "implicit demand response", "load scheduling", "markets", "price DR", "price-reactive systems", "rate-based", "uncoordinated" | 10, 28, 32, 61, 109–114 |
| Implicit cooperation | "autonomous control", "decentralised", "fully autonomous", "multi-agent" | 19, 30, 72, 95 |

**Table 5**
Summary of the detailed literature review of distributed energy resources coordination strategies mapped onto the taxonomy categories. For an extended version of this table, including a breakdown of each strategy, see Table B.1. Single citations may be listed under multiple categories if multiple strategies are defined in a paper. Each coordination category may be labelled by a variety of terms, while individual terms may be used to describe strategies across different categories. This reflects both the variety of possible strategies within each category and the lack of specific terminology, which leads to ambiguity.

# 4. Discussion: application of the taxonomy for coordination strategy selection

In this paper, we have developed a taxonomy on which we have shown that one can map any coordination strategy using non-ambiguous, objective structural classification criteria.

In this section, we illustrate the application of this taxonomy to select adequate types of coordination strategies based on the context of their application. We argue that to achieve deep decarbonisation, an ecosystem of control architectures suited to each specific context in all layers of the energy system is needed. Particularly, we highlight the need to identify and focus on under-integrated and under-researched niches. We then illustrate this argument using the taxonomy to highlight a potentially promising approach for coordinating residential energy, whose flexibility is so far under-utilised.

## 4.1. Selecting complementary coordination strategies for deep decarbonisation

Deep decarbonisation requires the coordination of a heterogeneous system of interlinked technologies, infrastructures, markets, regulations and user practices, such that every carbon-emitting activity contributes to efforts, beyond the low-hanging fruits [117]. In the context of energy systems, it involves unlocking the contributions to environmental and social welfare from assets at all levels, from the traditional centralised generation plant at transmission level to the heterogeneous distribution-network flexible loads beyond the reach of centralised strategies that have been traditionally used. Energy systems involve a plurality of stakeholders (e.g. business model developers, smart home systems manufacturers, individual generators and consumers, regulators, flexible loads aggregators) deploying a wide range of technologies that require control (e.g. microgrids, residential energy, industry, buildings, distributed renewable generation, transport fleet charging).

To identify appropriate DER coordination strategies for each specific context, researchers can borrow established heuristics to investigate how to maximise impact with limited resources. An obvious question to ask is “How many people benefit, and by how much?” [118]. The extensive benefits of coordinating DERs in various segments of energy systems have been identified, helping the safe and cost-effective integration of large shares of renewables to mitigate the current climate, health and environmental crises.

Further questions can then be asked for DER coordination strategy selection, starting with: “Is this the most effective thing you can do?” [118]. One size does not fit all; the choice of coordination category from the taxonomy in this paper will depend on individual coordination contexts and aims. Structurally differing coordination paradigms with specific advantages and drawbacks for different applications have been presented in Section 2. Identifying the type of strategy that is most cost-effective for each application is critical. For example, while it may be worth investing large amounts in communication and control infrastructure to fully exploit the flexibility potential of larger industrial assets, obtaining complete access to information and control to perform real-time centralised optimisation may be impractical or not most effective where the value of flexibility is smaller in individual distributed assets. The trade-offs between the additional value enabled by communication and the costs of implementing it, both in terms of equipment needed and of consumer acceptance, need to be particularly appraised. Other important criteria to be considered in individual contexts include the level of intelligence and flexibility of individual units, the type of prosumer (residential, commercial, industrial, military), stakeholder involvement, underlying motivation and drivers, supporting infrastructure required, integration with existing tariff structure, legal requirements, unit physical features (location, ownership, size), network assets response time, devices and communication failures, security risks, electrical grid conditions and desired robustness of the system [9, 10, 18, 49]. Coordination strategy-specific risks such as those related to privacy and security are mitigated by using multiple types of interoperable strategies suited to different challenges. Therefore, a heterogeneous mix of coordination strategies across the space of possible coordination categories is needed to go from shallow utilisation to deep integration of each remaining niche whose flexibility is so far underutilised.

In terms of setting priorities for the development of coordination strategies, another important question may be asked: “Is this area neglected?” [118]. We have shown that research is increasing exponentially in the field. There is a need to frame research relative to existing scholarship to avoid replicating efforts and focus on under-explored areas. As certain resources and tools have historically been extensively researched, for example the central optimisation of large power plants and industrial users, there is value in identifying under-integrated and under-researched segments with locked-in flexibility and value.

## 4.2. Potential path forward for future research

We now consider the residential sector as a broadly recognised remaining niche with so far under-integrated flexibility. This example illustrates the use of the taxonomy to select adequate coordination strategies in a given context.

Thanks to both storage and DR capacity, residential sites have a pivotal role to play in helping facilitate the integration of renewable energy generation. They represent for example 55.2% of the UK current total energy consumption for electricity, transport and heat, with the latter two undergoing electrification [119]. Increasing ownership of EVs and PV panels has been facilitated by regulation changes, with many countries phasing out internal combustion engine cars in the near future, and by plummeting costs, with an 89% and 85% levelised cost drop between 2010 and 2020 for EVs and PV panels and prospects of further declines in coming years [120, 121]. These represent significant untapped opportunities, as the provision of DR currently primarily focuses on industrial and commercial consumers [122], with most customers still limited to trade with utility companies [93].

Identifying the specific remaining hurdles for the coordination of residential energy flexibility, suitable coordination strategies can be identified using the taxonomy proposed in this paper. Firstly, computational and acceptance limitations mean that direct control would not be best suited at scale. There may be difficulties in computing central solutions due to scalability issues for large numbers of independent units, especially as residential customers are not motivated to invest substantial amounts in managing their electricity usage. Moreover, accurately forecasting individual residential consumption and generation profiles for day-ahead optimisation is challenging [10]. Secondly, due to both ICT infrastructure cost constraints as the residential flexibility potential is broken down into many small households with potential benefits too low to justify the cost of communication links, and to privacy concerns, personal information may not be readily shared in real time.

Implicit cooperation, which keeps personal information at the local level while encouraging cooperation towards global objectives, is an under-researched cooperation category that would be particularly suited to unlocking the potential systemic value of residential energy flexibility. As the communication and computation burden can become unwieldy in large-scale systems, pre-learned policies computed using RL and implemented in a decentralised fashion may help coordinate the so-far largely untapped residential energy flexibility. It avoids issues of costs, privacy and technical risks associated with both centralised and bilateral communication by allowing consumers to cooperate without sharing their data. Incentives could be designed to incorporate social objectives in the operation of distributed home energy systems without excessive interference in personal comfort and utility.

## 5. Conclusion

This paper has shown that the strategies for coordinating grid-edge energy resources could be classified in an exhaustive, mutually exclusive manner in a taxonomy based on the agency of controlled units, the structure of information flows, and the type of game agents play. The proposed taxonomy provides a clear and objective tool for transparently laying out structural assumptions in the design of future grid-edge resources coordination strategies.

We analysed the literature through the lens of this taxonomy with both a systematic literature review and a detailed mapping of 93 strategies onto the coordination categories. As fundamental changes in the make-up and the structure of the electricity grid are calling for transitions to new coordination paradigms, the systematic literature review identified fast emerging research themes such as the decentralisation of energy resources, growing privacy concerns, and distributed network constraints management.

As research efforts are increasing exponentially, doubling approximately every 4.7 years, we have highlighted using our novel framework how semantic clarity and effectiveness was at times hindered by the ambiguous terminology used in the field, with terms such as “multi-agent”, “peer-to-peer”, and “transactive energy” which may refer to structurally different coordination strategies. The detailed mapping of terminology and coordination strategies onto the coordination categories has illustrated the wealth coordination strategies and clarified their structural similarities and differences.

Clearly selecting the adequate control, communication and incentive types is key to incentivising the right actions to meet global goals. We have argued that a plurality of structurally different complementary coordination paradigms suited to each specific context in all layers of our electricity systems will help drive deep decarbonisation. Residential energy flexibility was identified as an under-integrated part of the energy system which may provide future opportunities. Interoperability between the different complementary coordination strategies within energy systems is another critical area for further research.

## Acknowledgement

This work was supported by the Saven European Scholarship and by the UK Research and Innovation and the Engineering and Physical Sciences Research Council (award references EP/S000887/1, EP/S031901/1, and EP/T028564/1).

## Contributions

F.C.: Conceptualization, Data curation, Formal Analysis, Investigation, Methodology, Visualization, Validation, Writing – original draft, Writing – review & editing
T.M.: Supervision, Validation, Writing – review & editing
M.M.: Supervision, Validation, Writing – review & editing

## Declaration of interests

The authors declare no competing interests.

| Type | Key words |
|---|---|
| Title or abstract | cloud, gas, forestry, data center, optical, water supply, mechanics, life cycle assessment, Micro Phasor Measurement, tyre, manufacturing, clutchless, concrete, methane, satellite, torque, health, speech |
| Journal | chemistry, social science, materials, environment, sustainable production and consumption, Vibration and Acoustics, nano energy, Future Generation Computer Systems, chemical engineering, food, surface science, Minerals Engineering |
| Subject Area | Medicine, Social sciences Arts and humanities, Environmental science, Mathematics, Agricultural and biological sciences, Physics and astronomy, Earth and planetary sciences, Biochemistry, genetics and molecular biology, Materials science, Psychology, Chemistry, Chemical engineering, Immunology and biology, Pharmacology, toxicology and pharmaceutics, Neuroscience, Nursing, Health professions, Veterinary, Dentistry |

**Table A.1**
Words excluded from topic search.

# A. Systematic literature review

## A.1. Scopus search query

We systematically review the literature using a structured topic search query in the Scopus search engine [33], the largest abstract and citation database of peer-reviewed literature. We aim to select literature that lies at the intersection of the concepts of coordination, grid-edge participation and electric resources. The following sequence is followed to define the query terms for the literature search:

1. Select terms associated with each of the concepts of coordination, grid-edge participation and electric resources. Where relevant, only the root of each term is selected so that associated adjectives, nouns and verbal forms may also be included in the search.
2. Conduct a search to obtain all publications containing at least one of the terms associated with each category in their title or abstract.
3. Inspect the titles and abstracts of the first 50 results, adding missing terms encountered that are relevant to any of the three categories, and excluding irrelevant search terms, topic areas and journals from future searches.
4. Repeat (2) and (3) until the first 50 results are either relevant or do not contain terms that may be excluded from the search without excluding more relevant results.

The resulting query terms are presented in Figure A.1. Terms which were excluded from the corpus are listed in Table A.1. The body of literature obtained comprised of publications published between 1962 to 2020. Inspecting the number of publications over time in Figure 1, we further focus our literature review on the period between 1995, the year in which the body of literature started growing consistently, and 2020, resulting in 73,053 titles and abstracts. As expected given current electric grid transformations, it is immediately apparent that the coordination of distributed grid-edge electric resources is an increasingly important topic of research. Publications including selected search themes rose exponentially from 248 in 1995 to 9,156 in 2020, doubling approximately every 4.7 years.

## A.2. Themes identification

Next, we identify research themes within this body of literature. Key words are identified from the first 50 titles and abstracts, and grouped in themes as listed in Table A.1. The absolute and relative prevalence of these major themes are displayed on Figure 3, so that trends may be identified. While the prevalence of themes in absolute numbers is highly dependent on the boundary of the body of literature defined in Appendix A.1, we identify research trends in relative terms by looking at prevalence over time, i.e. the share of publications within the body of literature that refers to each theme a given year. Moderate, fast and very fast increases correspond to average prevalences in the first half of the period (1995-2007) of less than 100%, two thirds and one third of that in 2020 respectively. In the rest of this section, we will individually describe the 17 major themes identified listed in Figure 3.

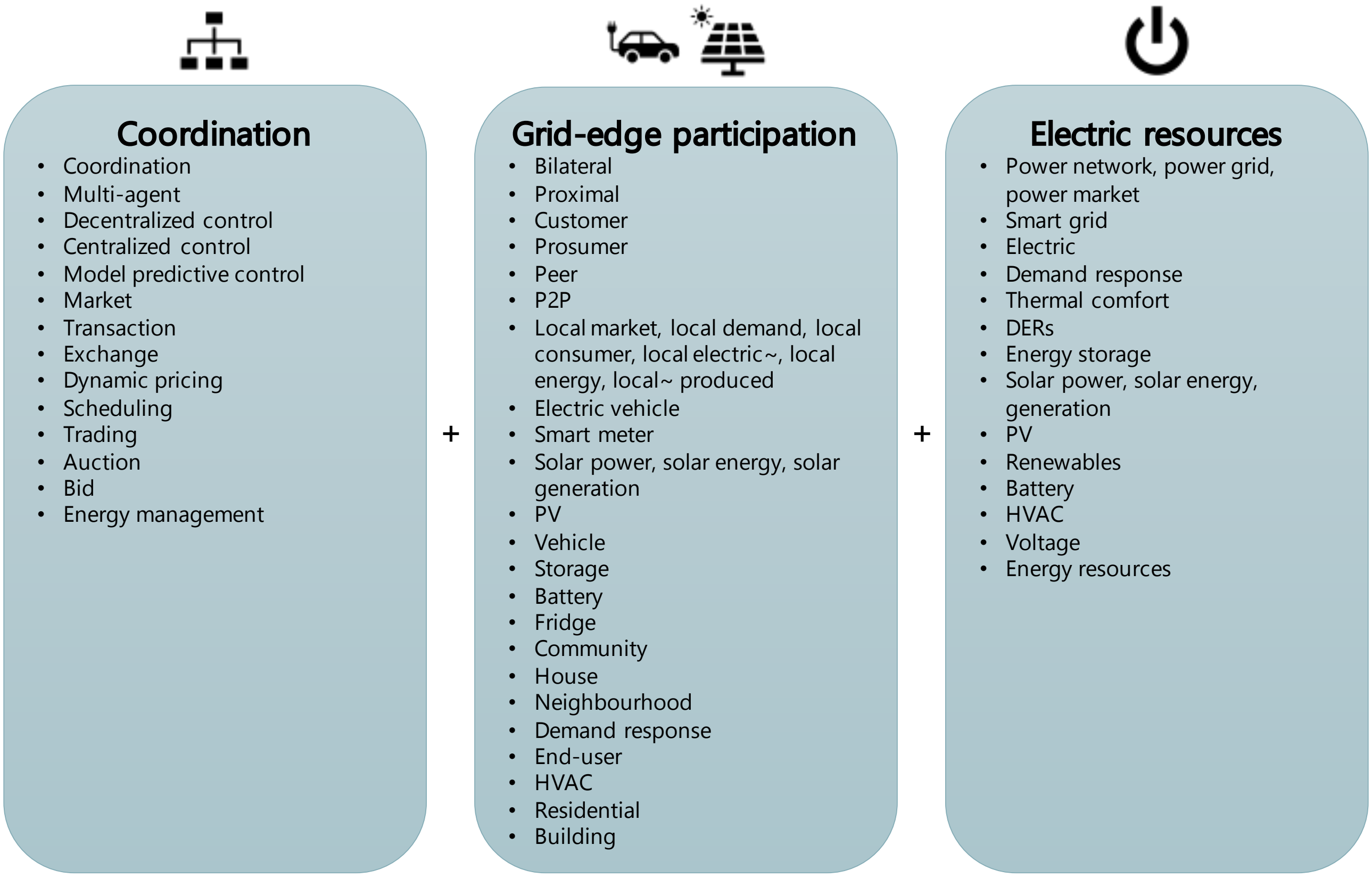

**Figure A.1:** Terms for the structured search query, associated with the concepts of coordination, grid-edge participation and energy resources.

| | Theme | Key words |
|---|---|---|
| 1 | Storage | storage, battery |
| 2 | Renewables | renewables, photovoltaic, PV, wind, hydro, solar power, solar generation, solar energy |
| 3 | Optimisation | optimise, maximise, minimise |
| 4 | Market | market, pricing, transaction, incentives |
| 5 | Network constraints | constraints, voltage, frequency, line capacity, stability, ancillary services, utility services |
| 6 | Uncertainty | uncertainty, risk, robust, stochasticity, forecasting, probabilistic |
| 7 | Demand Response | demand response, DR, demand side, customer side, flexible loads, flexibility, deferrable, load shifting |
| 8 | Residential sector | resident, domestic, house, household, home, community, neighbourhood |
| 9 | Electric vehicles | electric vehicle, EV |
| 10 | Decentralisation | decentralised, distributed |
| 11 | Thermal loads | thermal, temperature, heating, boiler, cooling, fridge, air conditioning, HVAC |
| 12 | Machine Learning | machine learning, ML, reinforcement learning, RL, artificial intelligence, AI, data-driven, neural network |
| 13 | Forecasting | forecast |
| 14 | Peer-to-peer | peer-to-peer, P2P, bilateral, matching |
| 15 | Game theory | game theory, game-theoretic, cooperation, coalition |
| 16 | Privacy | privacy, data protection, personal data, attack |
| 17 | Blockchain | blockchain, distributed ledger, DLT |

Table A.1: Research themes emerging from the keyword search.

## B. Strategies mapped onto taxonomy

In this section we build up on Table 5 to list and analyse individual strategies that have been mapped onto the taxonomy. Note that individual citations may be listed multiple times if multiple strategies were defined in a paper. Quoted descriptions for each strategy are included, which show that the labelling of the strategy in the paper alone does not allow the reader to unambiguously understand which paradigm it falls under. Where specified, we list individual units that are coordinated in the papers (local generation, storage, thermal control, generic flexible load), which shows that the unit that is coordinated is independent of the strategy classification which is agnostic to its application. The mark for strategies which use reinforcement learning also show that this tool my be used under various paradigms, as each paradigm may be implemented in numerous ways beyond the control, communication and incentive structure. We mark strategies which consider network and losses. Finally, the explicit objectives pursued by each strategy are categorised.

| Paradigm | Ref. | Strategy | Unit type(s) | | | | Methods | | | Objectives | | | | | | | |
|---|---|---|---|---|---|---|---|---|---|---|---|---|---|---|---|---|---|
| | | | | | | | | | | Minimise | | | | | Maximise | | |
| | | | Local generation | Storage | Thermal control | Generic flexible load | RL use | Network constraints | Networks losses | Operation costs | Investment costs | Battery degradation | Renewables curtailment | Power losses | Energy arbitrage, peak shaving | Load shifting, ancillary services | Prosumer utility |
| Direct control | 10 | "event-based DR" | | | | | | | | | | | | | | | |
| | 19 | "centralised control strategy" | | ✓ | | | | | | | | | | | | | |
| | 32 | "optimal power flow" | ✓ | ✓ | | | | ✓ | | | | | | | | | |
| | 32 | "direct load control" | ✓ | ✓ | | | | ✓ | | | | | | | | | |
| | 30 | "centralised control" | | | | | | | | | | | | | | | |
| | 46 | "multi-agent system" | ✓ | ✓ | | ✓ | | | | | | | | | ✓ | | |
| | 50 | "top-down switching" | | | | | | | | | | | | | | | |
| | 50 | "centralized optimization" | | | | | | | | | | | | | | | |
| | 61 | "operator instructions" | | | | | | | | | | | | | | | |
| | 70 | "centralised dispatch" | ✓ | ✓ | | ✓ | | ✓ | ✓ | | | | | | | | |
| | 78 | "model predictive control" | ✓ | ✓ | | | | ✓ | ✓ | | | | | ✓ | | | |
| | 79 | "stochastic optimization framework" | | | | | | ✓ | | | | | | | ✓ | ✓ | |
| | 80 | "model predictive control" | ✓ | ✓ | ✓ | ✓ | | ✓ | ✓ | ✓ | | | ✓ | | ✓ | | |
| | 81 | "centralized model predictive control scheme" | | ✓ | | | | ✓ | ✓ | | | ✓ | ✓ | | | | |
| | 82 | "optimal energy balance methodology" | ✓ | ✓ | ✓ | | | | | ✓ | ✓ | | | | | | |
| | 83 | "consumer automated energy management system" | ✓ | | | ✓ | ✓ | | | ✓ | | | | | | | ✓ |
| | 84 | "fridge controller" | | | ✓ | | | | | | | | | | | ✓ | |
| | 85 | "direct load control architecture" | | | ✓ | | | | | | | | | | | ✓ | |
| Mediated competition | 10 | "organized markets" | | | | | | | | | | | | | | | |
| | 10 | "demand reduction bids" | | | | | | | | | | | | | | | |

| Paradigm | Ref. | Strategy | Unit type(s): Local generation | Unit type(s): Storage | Unit type(s): Thermal control | Unit type(s): Generic flexible load | Methods: RL use | Methods: Network constraints | Methods: Networks losses | Objectives, Minimise: Operation costs | Objectives, Minimise: Investment costs | Objectives, Minimise: Battery degradation | Objectives, Minimise: Renewables curtailment | Objectives, Minimise: Power losses | Objectives, Maximise: Energy arbitrage, peak shaving | Objectives, Maximise: Load shifting, ancillary services | Objectives, Maximise: Prosumer utility |
|---|---|---|---|---|---|---|---|---|---|---|---|---|---|---|---|---|---|
| | 16 | “adaptive consumption-level pricing scheme” | | | | | | | | | | | | | | | ✓ |
| | 26 | “coordinated market” | | | | | | | | ✓ | | | | | | | ✓ |
| | 26 | “community market” | | | | | | | | | | | | | | | |
| | 28 | “explicit demand response” | | | ✓ | | | | | | | | | | | ✓ | |
| | 32 | “demand response” | ✓ | ✓ | | | | ✓ | | | | | | | | | |
| | 32 | “auction-based methods” | ✓ | ✓ | | | | ✓ | | | | | | | | | |
| | 49 | “transactive energy service system” | | | | | | | | | | | ✓ | | | | |
| | 50 | “transactive control” | | | | | | | | | | | | | | | ✓ |
| | 51 | “market-based multi-agent system” | | ✓ | | ✓ | ✓ | | | | | | | | ✓ | | |
| | 52 | “market-based control” | | ✓ | | | ✓ | | | ✓ | | | | | | | ✓ |
| | 54 | “market-based EV charging coordination” | | ✓ | | | ✓ | | | | | | | | | | ✓ |
| | 55 | “reinforcement learning-based dynamic pricing algorithm” | | | | ✓ | ✓ | | | | | | | | | | ✓ |
| | 56 | “system-centric” | | | | | | ✓ | ✓ | ✓ | | | | | | | ✓ |
| | 57 | “continuous double auction” | ✓ | ✓ | | | | ✓ | ✓ | ✓ | | | | ✓ | ✓ | ✓ | ✓ |
| | 58 | “P2P local electricity market model” | | | ✓ | ✓ | | ✓ | | ✓ | | | | | | ✓ | |
| | 59 | “energy blockchain in microgrids” | ✓ | | | | | ✓ | ✓ | | | | | | | ✓ | ✓ |
| | 60 | “decentralized markets for distribution system flexibility” | ✓ | ✓ | | ✓ | | ✓ | ✓ | | | | | | | ✓ | ✓ |
| | 61 | “price-responsive modes” | | | | | | | | | | | | | | | |
| | 67 | “P2P trading” | ✓ | | | | | | | ✓ | | | ✓ | | | | ✓ |

| Paradigm | Ref. | Strategy | Unit type(s): Local generation | Unit type(s): Storage | Unit type(s): Thermal control | Unit type(s): Generic flexible load | Methods: RL use | Methods: Network constraints | Methods: Networks losses | Objectives, Minimise: Operation costs | Objectives, Minimise: Investment costs | Objectives, Minimise: Battery degradation | Objectives, Minimise: Renewables curtailment | Objectives, Minimise: Power losses | Objectives, Maximise: Energy arbitrage, peak shaving | Objectives, Maximise: Load shifting, ancillary services | Objectives, Maximise: Prosumer utility |
|---|---|---|---|---|---|---|---|---|---|---|---|---|---|---|---|---|---|
| | 70 | “unidirectional pricing” | ✓ | ✓ | | ✓ | | ✓ | ✓ | ✓ | | ✓ | | ✓ | | | |
| | 75 | “automatic P2P energy trading model based on reinforcement learning” | ✓ | ✓ | | | ✓ | | | | | | ✓ | | | | |
| | 86 | “incentive-based demand response” | | | | ✓ | ✓ | | | | | | | | | ✓ | ✓ |
| | 87 | “agile demand response” | | | | ✓ | ✓ | | | ✓ | | | | | | | ✓ |
| | 88 | “Stackelberg approach” | | | | ✓ | | | | | | | | | | | ✓ |
| | 89 | “distributed demand response” | | | | ✓ | ✓ | | | ✓ | | | | | ✓ | | ✓ |
| | 90 | “bilateral trading electricity market” | ✓ | | | ✓ | | | | | | | | | | ✓ | ✓ |
| | 91 | “deep reinforcement learning for strategic bidding” | ✓ | | | ✓ | ✓ | | | | | | | | | | ✓ |
| | 92 | “learning based bidding strategy” | | | ✓ | | ✓ | | | | | | | | | | ✓ |
| | 93 | “indirect customer-to-customer energy trading” | | | | | ✓ | | | ✓ | | | | | ✓ | ✓ | ✓ |
| Mediated cooperation | 11 | “energy collectives” | ✓ | | | ✓ | | | | ✓ | | | | | | | ✓ |
| | 40 | “prosumer coalitions with energy management” | | ✓ | | ✓ | | | | | | | | | | | ✓ |
| | 47 | “community-based market” | | | | | | | | | | | | | | | |
| | 53 | “deep transfer Q-learning” | ✓ | | | ✓ | ✓ | | | ✓ | | | | | | | ✓ |
| | 62 | “multi-agent residential demand response” | | ✓ | | | ✓ | | | | | | | | ✓ | | |
| | 63 | “decentralized learning-based multi-agent residential DR” | | ✓ | | | ✓ | | | | | | ✓ | | ✓ | | |
| | 64 | “extended joint action learning” | | | ✓ | ✓ | ✓ | ✓ | | | | | | | | ✓ | ✓ |
| | 65 | “P2P trading” | | | | | | | | | | | | | ✓ | | ✓ |
| | 73 | “distributed price-directed optimization” | ✓ | ✓ | | | | | | ✓ | | ✓ | | ✓ | | | ✓ |

| Paradigm | Ref. | Strategy | Unit type(s): Local generation | Unit type(s): Storage | Unit type(s): Thermal control | Unit type(s): Generic flexible load | Methods: RL use | Methods: Network constraints | Methods: Networks losses | Objectives – Minimise: Operation costs | Objectives – Minimise: Investment costs | Objectives – Minimise: Battery degradation | Objectives – Minimise: Renewables curtailment | Objectives – Minimise: Power losses | Objectives – Maximise: Energy arbitrage, peak shaving | Objectives – Maximise: Load shifting, ancillary services | Objectives – Maximise: Prosumer utility |
|---|---|---|---|---|---|---|---|---|---|---|---|---|---|---|---|---|---|
| | 95 | "semiautonomous operation" | | | ✓ | | | | | | | | | | ✓ | ✓ | |
| | 96 | "coordinated multilateral trades" | | | | | | ✓ | ✓ | | | | | | | ✓ | ✓ |
| | 97 | "distribution locational marginal costs and hierarchical decomposition" | ✓ | ✓ | | | | | | ✓ | | | | | | ✓ | |
| Bilateral competition | 26 | "decentralized market" | | | | | | | | | | | | | | | |
| | 32 | "matching theory-based methods" for "P2P" | ✓ | ✓ | | | | ✓ | | | | | | | | | |
| | 56 | "peer-centric" | | | | | | | | | | | | | | | ✓ |
| | 61 | "bilateral transactive bids" | | | | | | | | | | | | | | | |
| | 66 | "prosumer centric local energy market" | | | | | | | | | | | | | | | ✓ |
| | 67 | "P2P trading" | ✓ | | | | | | | | | | | | | | |
| | 68 | "blockchain-based distributed double auction trade" | ✓ | | | | | ✓ | ✓ | | | | | | | | ✓ |
| | 69 | "bilateral contract networks" | ✓ | ✓ | | ✓ | | | | | | | | | | | ✓ |
| | 70 | "P2P energy trading" | ✓ | ✓ | | ✓ | | | | | | | | | | | |
| | 106 | "bilateral contracts" | | | | | | | | | | | | | | | ✓ |
| Bilateral cooperation | 19 | "distributed multi-agent control strategy" | | ✓ | | | | | | | | | | | | | |
| | 30 | "distributed control" | | | | | | | | | | | | | | | |
| | 32 | "distributed methods" | ✓ | ✓ | | | | ✓ | | | | | | | | | |
| | 47 | "full P2P market" | | | | | | | | ✓ | | | | | | | |
| | 48 | "decentralized bilateral energy trading" | | | | | | ✓ | | | | | | | | | ✓ |
| | 70 | "distributed dispatch" | ✓ | ✓ | | ✓ | | | | | | | | | | | |

| | | | Unit type(s) | | | | Methods | | | Objectives | | | | | | | | |
|---|---|---|---|---|---|---|---|---|---|---|---|---|---|---|---|---|---|---|
| | | | | | | | | | | Minimise | | | | | | Maximise | | |
| Paradigm | Ref. | Strategy | Local generation | Storage | Thermal control | Generic flexible load | RL use | Network constraints | Networks losses | Operation costs | Investment costs | Battery degradation | Renewables curtailment | Power losses | Energy arbitrage, peak shaving | Load shifting, ancillary services | Prosumer utility |
| | 71 | "parallel transfer learning" | | ✓ | | | ✓ | | | | | | | | ✓ | ✓ | |
| Implicit competition | 10 | "rate-based or price DR programs" | | | | | | | | | | | | | | | |
| | 28 | "implicit demand response" | | | ✓ | | | | | | | | | | ✓ | | ✓ |
| | 32 | "uncoordinated approaches" | ✓ | ✓ | | | | ✓ | | | | | | | | | |
| | 50 | "price-reactive systems" | | | | | | | | | | | | | | | |
| | 61 | "autonomous mode" | | | | | | | | | | | | | | | ✓ |
| | 109 | "load scheduling algorithm" | | | | ✓ | | | | | | | | | | | ✓ |
| | 110 | "deep reinforcement learning based energy storage arbitrage" | | ✓ | | | ✓ | | | | | ✓ | | | | | ✓ |
| | 111 | "energy storage arbitrage in real-time markets via teinforcement learning" | | ✓ | | | ✓ | | | | | | | | | | ✓ |
| | 112 | "device-based reinforcement learning" | | | | ✓ | ✓ | | | | | | | | | | ✓ |
| | 113 | "reinforcement learning controller" | | | ✓ | | ✓ | | | | | | | | | | ✓ |
| | 114 | "reinforcement learning control" | ✓ | | ✓ | | ✓ | | | | | | ✓ | | | | ✓ |
| Implicit cooperation | 19 | "decentralised control strategy" | | ✓ | | | | | | | | | | | | ✓ | |
| | 30 | "autonomous control" | | | | | | | | | | | | | | ✓ | |
| | 72 | "prediction-based multi-agent reinforcement learning" | | ✓ | | | ✓ | | | | | | | | ✓ | | |
| | 95 | "fully autonomous operation" | | | ✓ | | | | | | | | | | | ✓ | |

| Paradigm | Ref. | Strategy | Unit type(s) | | | | Methods | | | Objectives | | | | | | | | |
|---|---|---|---|---|---|---|---|---|---|---|---|---|---|---|---|---|---|---|
| | | | | | | | | | | Minimise | | | | | Maximise | | | |
| | | | Local generation | Storage | Thermal control | Generic flexible load | RL use | Network constraints | Networks losses | Operation costs | Investment costs | Battery degradation | Renewables curtailment | Power losses | Energy arbitrage, peak shaving | Load shifting, ancillary services | Prosumer utility | |

Table B.1: Distributed energy resources coordination strategies mapped onto the proposed taxonomy. Ticks indicate which controlled resources are specifically modelled (local generation, storage, thermal control, generic flexible load), whether reinforcement learning (RL) is being used as an example of coordination tool, and whether the impact of coordination strategies on network constraints and losses is modelled. In this table energy arbitrage may refer to displacing energy use to times of lower prices or to times with higher renewables generation. Ancillary services may refer to provision of reactive power and real-time management of network constraints including voltage and frequency control. No ticks are indicated were an individual article does not specify the coordinated unit type or specific objective.